\documentclass{iopart}
\usepackage{iopams}
\usepackage{graphicx,latexsym}

\def\bu{{\bf u}}
\def\ba{{\bf a}}
\def\cA{{\cal A}}

\def\rmi{\mathrm{i}}
\def\openone{\leavevmode\hbox{\small1\kern-3.5pt\normalsize1}}

\def\bbbn{{\Bbb N}}
\def\bbbc{{\Bbb C}}
\def\bbbr{{\Bbb R}}

\def\bbbs{{\Bbb S}}
\def\bbbp{{\Bbb P}}

\def\re{\mbox{Re\,}}
\def\diag{\mbox{diag\,}}

\newtheorem{remark}{Remark}

\begin{document}

\title{Reductions of integrable equations on  {\bf A.III}-type symmetric spaces}

\author{V S Gerdjikov$^1$, A V Mikhailov$^2$, and T I Valchev$^1$}

\address{$^1$ Institute for Nuclear Research and Nuclear Energy,\\
Bulgarian Academy of Sciences, 72 Tsarigradsko chaussee,
1784 Sofia, BULGARIA}

\address{$^2$ Applied Math. Department, University of Leeds, \\Leeds, LS2 9JT, UK
}

\eads{\mailto{gerjikov@inrne.bas.bg}, \mailto{a.v.mikhailov@leeds.ac.uk}
, \mailto{valtchev@inrne.bas.bg}}

\begin{abstract}
We study a class of integrable non-linear differential equations related to the {\bf A.III}-type
symmetric spaces. These spaces are realized as factor groups of the form
$SU(N)/S(U(N-k)\times U(k))$. We use the Cartan involution corresponding to this symmetric space as an element of the reduction group
and restrict generic Lax operators to this symmetric space. The symmetries of the Lax operator are inherited by the fundamental analytic
solutions  and give a characterization of the corresponding Riemann-Hilbert data.
\end{abstract}

\noindent{\it Keywords\/}: integrable equations, symmetric spaces, Riemann-Hilbert problem

\pacs{02.20.Sv, 02.30.Ik, 02.30.Zz}
\submitto{\JPA}

\maketitle

\section{Introduction}
Systems of nonlinear partial differential equations, integrable by the inverse transform method, can be obtained as reductions of
generic integrable systems corresponding to  Lax operators with matrix coefficients. For example, the coefficients of
a generic Lax operators are elements of the Lie algebra $sl(N,\bbbc)$ and simplest reductions are just restrictions on (simple)
subalgebras of $\cA\subset sl(N,\bbbc)$. Restrictions on Kac-Moody subalgebras of the loop algebras
$\bbbc[\lambda,\lambda^{-1}]\otimes_{\bbbc}\cA$ lead to interesting classes of
integrable equations \cite{DrSok*85,mik_toda,mik,mop1}. Further generalizations give rise to the concept of
automorphic Lie algebras, which are subalgebras of   $\cA(\lambda)=\bbbc(\lambda)\otimes_{\bbbc}\cA$  \cite{mik,miklom,lomsan}.
In the latter approach the subgroups of the group of automorphisms of the loop algebra or more general of the algebra  $\cA(\lambda)$ play the central r\^ole.
In the context of the reduction problem these subgroups (the reduction groups) were introduced and studied in
\cite{mik_toda,miktetr,mik,miklom1,miklom}.
In order to apply the inverse spectral transform to the reduced equations one needs to give a characterization of the reduction
in terms of the spectral data. The reduction group naturally acts on the analytic fundamental solutions of the linear problem
corresponding to the Lax operator, on the scattering and Riemann-Hilbert data. Continuous and discrete spectrum of the operator
are orbits of the reduction group \cite{mik,mik_ll}.

According to Cartan (see \cite{Hel}) the local structure of a symmetric space is determined by an
involutive automorphism $\varphi_1$ of the relevant Lie algebra $\mathfrak{g} $, known as Cartan involution,
and the corresponding decomposition
$\mathfrak{g} = \mathfrak{g} ^{(0)}\oplus\mathfrak{g} ^{(1)}, \
\mathfrak{g} ^{(n)} = \{ a\in \mathfrak{g}\,|\,  \varphi_1(a) = (-1)^n a\}.$
In this paper we begin with the algebra $sl(N,\bbbc)$ and use the automorphism $\varphi_*(a)=-a^\dag$ to reduce
$sl(N,\bbbc)$ to $\varphi_*$--invariant subalgebra   $\mathfrak{g}
=\{a\in sl(N,\bbbc)\,|\, \varphi_*(a)=a\}$, thus $\mathfrak{g}=su(N)$. Choosing the Cartan involution of the form
$\varphi_1(a)={\bf J}_ka{\bf J}_k$ with ${\bf J}_k=\diag(1, \dots, 1, -1,\dots, -1) $ we obtain the decomposition of $su(N)$
which reflects the local structure of the compact {\bf A.III}-type symmetric space $SU(N)/S(U(k)\times U(N-k))$.
The automorphisms $\varphi_*,\varphi_1$  can be extended to the loop algebra
$\cA_{\lambda}=\bbbc[\lambda,\lambda^{-1}]\otimes_{\bbbc}sl(N,\bbbc)$ as $\Phi_*(a(\lambda))=-a^\dag(\lambda^*)$ and
$\Phi_1(a(\lambda))=\varphi_1(a(-\lambda))$.

In Section 2 we analyze Lax operators that are linear in $\lambda$ and invariant with respect to the reduction group generated by the
automorphisms $\Phi_*(a(\lambda)),\Phi_1(a(\lambda))$.
They give rise to the integrable system
\begin{equation}
 \label{eqiso_Int}
i\bu_t=((1-\bu\bu^{\dag})\bu_x)_x\, , \qquad \bu^{\dag} \bu=\openone _k\, ,
\end{equation}
where $\bu$ is $(N-k)\times k$ complex matrix and $\openone _k$ is a unit matrix. System (\ref{eqiso_Int}) is $S(U(N-k)\times U(k))$ invariant
and in this sense isotropic. In particular, if $k=1$ equation (\ref{eqiso_Int}) can be seen
as a $U(N-1)$ invariant integrable system on $\bbbc\bbbp^{N-1}$. In Section 2 we discuss this reduction in details.

The loop algebra $\cA_{\lambda}$ has automorphisms of the form $\Phi_2(a(\lambda))=J_2a(\lambda^{-1})J_2^{-1}$. The simplest Lax operator which is
invariant with respect to automorphisms $\Phi_*,\Phi_1$ and $\Phi_2$ is a ``symmetric'' Laurent polynomial and has simple poles in $\lambda$
at points $\{0,\infty\}$. In Section 3 we study Lax operators and integrable equations related to such operators.
The simplest non-trivial system of this type is of the form:
\begin{eqnarray}\label{eq:u-v1}
iu_t&=&u_{xx}-(u(u^*u_x+v^*v_x))_x+8 vv^*u,\\ \label{eq:u-v2}
iv_t&=&v_{xx}-(v(u^*u_x+v^*v_x))_x-8uu^*v\, ,
\end{eqnarray}
where $u$ and $v$ are functions of $x$ and $t$ subject to the condition:
\begin{equation}\label{eq:uv0}
|u^2| + |v^2| = 1
\end{equation}
i.e. the vector with components $u$ and $v$ sweeps a $3$-dimensional
sphere in $\bbbr^4$. System (\ref{eq:u-v1}),  (\ref{eq:u-v2}) can also be seen as an anisotropic
deformation of (\ref{eqiso_Int}) with $k=1,N=3$.

In Section 4 we formulate the spectral properties of the Lax operator on the class of
potentials satisfying (\ref{eq:uv0}).  We outline the construction of the fundamental analytic solutions of $L$. As a result
we are able to reduce the inverse scattering problem  for $L$ to a Riemann-Hilbert problem.
This Riemann-Hilbert problem does not allow canonical normalization due to the nontrivial asymptotics of fundamental analytic solutions
both for $\lambda\to\infty$ and $\lambda\to 0$ -- the two singular points. The canonical normalization can be partially replaced by the requirement of
the invariance of the fundamental analytic
solutions with respect to the reduction group (see \cite{mik_ll}).

In Section 5 we analyze the mapping $\mathcal{F}$ between the potential and the scattering data of $L$.
Using the Wronskian relations we  introduce the `squared solutions'. Thus the mapping $\mathcal{F}$ can be interpreted as
a generalized Fourier transform \cite{GVY*08}.

\section{Lax representation. The isotropic case}

Let us consider two linear differential operators (the Lax representation) with $N\times N$ matrix coefficients
\begin{equation}\label{L_iso}
 L=D_x+i\left(\begin{array}{cc}
         {\bf 0}_k&\lambda {\bf u}^{\dag}\\
\lambda {\bf u}&{\bf 0}_{N-k}
        \end{array}
\right),
\end{equation}
\begin{equation}\label{A_iso}
 A=D_t+\left(\begin{array}{cc}
         i\lambda^2 {\bf u}^{\dag}{\bf u}&-\lambda {\bf a}^{\dag}\\
\lambda {\bf a}&i\lambda^2 {\bf u}{\bf u}^{\dag}
        \end{array}
\right),
\end{equation}
where $D_x,D_t$ are operators of differentiation, ${\bf 0}_s$ denotes a square $s\times s$ zero matrix, ${\bf u}$
and ${\bf a}$ are complex $(N-k) \times k$ matrices whose entries are differentiable functions of independent variables $(x,t)$, ${\bf u}^{\dag},{\bf a}^{\dag}$
denote Hermitian conjugated matrices and $\lambda$
is a {\sl spectral} parameter.

The commutativity condition $[L,A]=0$ is equivalent to the system of equations
\begin{eqnarray}
 &&\bu^{\dag}\ba+\ba^{\dag}\bu=-\bu^{\dag}_x\bu-\bu^{\dag}\bu_x\label{LAeq1}\\
&&\bu\ba^{\dag}+\ba\bu^{\dag}=\bu_x\bu^{\dag}+\bu\bu_x^{\dag}\label{LAeq2}\\
&& i\bu_t=\ba_x.\label{LAeq3}
\end{eqnarray}
It follows from (\ref{LAeq1}),(\ref{LAeq2}) that $(\mbox{tr} (\bu^{\dag}\bu)^n)_x=0,\ n\in\bbbn$,
thus the eigenvalues of the matrix $\bu^{\dag}\bu$ are functions of $t$ only.
In the case $k=1$, we have $\bu^{\dag} \bu=f(t)\in\bbbr$ and changing the variable $t\rightarrow T$, so that $D_t=f(t)D_T$, we can
set $\bu^{\dag} \bu=1$ without any loss of generality.
If $1<k<N-k$ then we shall assume that $\bu^{\dag} \bu=\openone_k$, where $\openone_k$ is the unit $k\times k$ matrix (this choice is consistent with
the system (\ref{LAeq1})-(\ref{LAeq3}), but is not the most general one). Then, it follows from  (\ref{LAeq1})-(\ref{LAeq2}) that
$$
 \ba= (\openone -\bu\bu^{\dag})\bu_x+i\gamma(x,t)\bu
$$
where $\gamma(x,t)\in\bbbr$ is an arbitrary real function which without loss of generality can be set to zero after an appropriate change of the coordinates
$t\rightarrow t,\ x\rightarrow X(x,t)$. Thus  we assume
\begin{equation}
 \label{bfa}
\ba= (\openone -\bu\bu^{\dag})\bu_x
\end{equation}
and equation (\ref{LAeq3}) takes the form
\begin{equation}
 \label{eqiso}
i\bu_t=((\openone -\bu\bu^{\dag})\bu_x)_x\, , \qquad \bu^{\dag} \bu=\openone _k\, .
\end{equation}
System (\ref{eqiso}) is $U(N-k)\times SU(k)$ invariant.
In the case $k=1,N\ge 3$ system (\ref{eqiso}) is $U(N-1)$ invariant vector equation and it can be seen as an
integrable system on $2N-3$ dimensional real sphere:
\begin{equation}
 \label{eqre}
\fl \eqalign{
\vec{u}_{t} &=\vec{v}_{xx}-(((\vec{u}\cdot \vec{v}_x)-(\vec{v}\cdot \vec{u}_x)) \vec{u})_x\, ,\\
-\vec{v}_{t}&=\vec{u}_{xx}+(((\vec{u}\cdot \vec{v}_x)-(\vec{v}\cdot \vec{u}_x)) \vec{v})_x\, ,}
\qquad (\vec{u}\cdot \vec{u})+(\vec{v}\cdot \vec{v})=1,\quad \vec{u}, \vec{v}\in\bbbr^{N-1}\, ,
\end{equation}
where $\vec{u}=\mbox{Re}\,\bu,\ \vec{v}=\mbox{Im}\,\bu$.

The Lax representation (\ref{L_iso}) can be naturally related to a reduction group generated by two automorphisms.
Let us consider general linear operators of the form
\begin{equation}\label{Lgen}
\fl  L=D_x+A_0+\lambda A_1 ,\qquad A=D_t+B_0+\lambda B_1 +\lambda^2 B_2
\end{equation}
where $A_0,A_1,B_0,B_1,B_2\in sl(N,\bbbc)$ are matrix functions of $x,t$. On the Lie
algebra ${\cal A}_\lambda=\bbbc[\lambda]\otimes sl(N,\bbbc)$ (i.e. the polynomial part of the corresponding loop algebra)
there is the outer automorphism $\Phi_*:{\cal A}_\lambda\mapsto{\cal A}_\lambda$ defined as $\Phi_*(a(\lambda))=-a^{\dag}(\lambda^*)$.
It is obvious that the $\Phi_*$ invariant subalgebra
$\{a(\lambda)\in{\cal A}_\lambda\,|\,\Phi_*(a(\lambda))=a(\lambda)\}=\bbbc[\lambda]\otimes su(N)$. Restriction to this subalgebra
is an obvious reduction of the general Lax pair (and the corresponding equations). In terms of operators $L,A$ this restriction
is equivalent to the condition
\begin{equation}
 \label{reds1}
L^{\dag}(\lambda^*)=L^{\rm ad}(\lambda),\qquad A^{\dag}(\lambda^*)=A^{\rm ad}(\lambda),
\end{equation}
where $L^{\dag}$ stands for the Hermitian conjugation of the operator $L$ and $L^{\rm ad}$ for the adjoint operator of $L$.

Further restriction can be achieved using the Cartan involutive  automorphism
$\varphi_1: su(N)\mapsto  su(N)$
which is defined as $\varphi_1(a)={\bf J}_k a {\bf J}_k$ where ${\bf J}_k=\mbox{diag}\,(j_1,\ldots ,j_N),\ j_1=\cdots
= j_k=1, \ j_{k+1}=\cdots =j_N=-1$ and we also assume $N-s>s$.
Automorphism $\varphi_1$ induces a grading in the algebra $\mathfrak{g}=su(N)$
\begin{equation}\label{eq:4}
\mathfrak{g} = \mathfrak{g} ^{(0)}\oplus\mathfrak{g} ^{(1)}, \qquad
\mathfrak{g} ^{(n)} = \{ a\in \mathfrak{g}\,|\,  {\bf J}_k a  {\bf J}_k^{-1} = (-1)^n a\}.
\end{equation}
The Cartan automorphism $\varphi_1$ can be extended to the automorphism of the corresponding loop algebra
$$\Phi_1:\bbbc[\lambda]\otimes su(N)\mapsto \bbbc[\lambda]\otimes su(N),\qquad \Phi_1(a(\lambda))={\bf J}_k a(-\lambda){\bf J}_k,$$

Operator $L$ (\ref{Lgen}) restricted on the subalgebra invariant with respect to the both automorphisms $\Phi_*$ and $\Phi_1$
satisfies the symmetry conditions (\ref{reds1}) and
\begin{equation}\label{reds2}
  {\bf J}_k L(-\lambda) {\bf J}_k^{-1}=L(\lambda)
\end{equation}
and is of the form
\begin{equation}
 \label{lnog}
L=D_x+i\left(\begin{array}{cc}
          R&\lambda \hat{{\bf u}}^{\dag} \\ \lambda \hat{{\bf u}}&Q
        \end{array}
\right),
\end{equation}
where $R,Q$ are Hermitian matrices of the size $k\times k$ and $(N-k)\times (N-k)$ respectively. By an appropriate  gauge transformation
$L\mapsto G^{\dag}LG, \ G\in S(U(k)\times U(N-k))$ we can set $R={\bf 0}_k$ and $Q={\bf 0}_{N-k}$. Then the operator $A$ can be found from the
condition $[L,A]=0$ as described above. The operator $A$ also satisfies the symmetry condition ${\bf J}_k A(-\lambda) {\bf J}_k^{-1}=A(\lambda)$.

Taking other simple Lie algebras and replacing the automorphism $\varphi_1$ by an appropriate Cartan automorphism the construction
described in this section can be easily extend our construction to other symmetric spaces.

\section{Lax representation. The anisotropic case}

The loop algebra $\hat{{\cal A}}_\lambda=\bbbc[\lambda,\lambda^{-1}]\otimes su(N) $ has a richer group of automorphisms than its
polynomial part $\bbbc[\lambda]\otimes su(N)$. Let us take an involutive automorphism $\varphi_2:su(N)\mapsto su(N),
\ \varphi_2\ne\varphi_1$.
In this paper  we shall assume that automorphism $\varphi_2$ is inner and of the form
\begin{equation}
 \varphi_2(a)=J_2aJ_2, \qquad J_2^2=I, \qquad [J_2,J_1]=0.
\end{equation}
Therefore it
commutes with the Cartan automorphism $\varphi_1$ discussed in the previous Section (more general construction of the reduction groups and corresponding automorphic Lie
algebras, including outer and non-commutative automorphisms are discussed in \cite{mik,miklom}). The map $\Phi_2: \hat{{\cal A}}_\lambda\mapsto \hat{{\cal A}}_\lambda$
defined as $$ \Phi_2(a(\lambda))=\varphi_2( a(\epsilon\lambda^{-1})),\quad \epsilon\ne 0,\ \epsilon\in\bbbr $$
is an automorphism of the loop algebra. In what follows we shall assume that the matrices $J_1, J_2$ are of the form
$$ J_1=\left(\begin{array}{cc}
              -1&0\\0& \openone
             \end{array}\right),\qquad
J_2=\left(\begin{array}{cc}
              1&0\\0&{\bf J}
             \end{array}\right), \qquad {\bf J}^2=\openone
$$
where $\openone $ is a unit $(N-1)\times (N-1)$ matrix.

The Lax representation with operators $L,A$ having simple and double poles in $\lambda$ respectively and
invariant with respect to the automorphisms $\Phi_*,\Phi_1$ and $\Phi_2$ can be written in the form
\begin{equation}
 \label{LA_anis}
\eqalign{ L&=D_x-i\left(\begin{array}{cc}
             0&\bu^\dag \Lambda \\ \Lambda\bu&0
             \end{array}\right),\\
A&=D_t+iA_0+i\left(\begin{array}{cc}
             -4\epsilon (\bu^\dag {\bf J}\bu)+\bu^\dag \Lambda^2 \bu& -i\ba^{\dag}\Lambda \\i\Lambda\ba&\Lambda\bu\bu^{\dag}\Lambda
             \end{array}\right)}
\end{equation}
where $\ba$ is given by (\ref{bfa}),  $\Lambda=\lambda {\bf I}+\epsilon \lambda^{-1}{\bf J}$ and
$A_0$ is a real constant  matrix of the form
$$ A_0=\left(\begin{array}{cc}
              0&0\\0&{\bf A}
             \end{array}\right),\qquad [{\bf A,J}]=0.
$$
The compatibility condition $[L,A]=0$ of the above operators (\ref{LA_anis})
leads to an anisotropic (deformation)
integrable system (\ref{eqiso}) with $k=1$\footnote{Here we have to note that when this paper had been completed, one of the authors
(AVM) contacted
V.V.Sokolov to discuss equation (\ref{eqaniso}). Sokolov draw our attention to his paper \cite{golsok}, where a similar equation and the
corresponding Lax representation had been found. His system was related to $sl(N,\bbbr)$ algebra (rather than to $su(N)$, as in our case)
and it was more general: in \cite{golsok} the matrix ${\bf J}$ is an arbitrary real matrix, without the condition ${\bf J}^2=\openone$.
In \cite {golsok} the Lax operators were not related to any reduction group. The reduction group reflects symmetries of the operator  which
are very useful for the spectral characterization of
the operators (see next sections of this paper).}:
\begin{equation}
 \label{eqaniso}
\fl \rmi \bu_t=(\bu_x-\bu(\bu^{\dag}\cdot \bu_x))_x+4\epsilon \bu (\bu^\dag\cdot{\bf J}\bu)+{\bf A}\bu
    \, , \qquad \bu^{\dag} \bu=1, \ \bu\in\bbbc^{N-1}\, .
\end{equation}

In the next Sections we shall develop the spectral theory for the operator $L$ in the simplest nontrivial case $N=3$.
We take automorphism $\varphi_1$ with $J_1=\mbox{\diag} (1,-1,-1)$, automorphism
 $\varphi_2$ with $J_2=\mbox{\diag} (1,-1,1)$ and set $\epsilon=1$. In this case the  Lax operator $L$ (\ref{LA_anis}) we shall write
the form
\begin{equation}
 \label{L_3}
L=iD_x+U(x,\lambda), \qquad U(x,\lambda)=\lambda L_1+\lambda^{-1} L_{-1}
\end{equation}
where
\begin{equation}
 \label{L1}
 L_1 = \left(\begin{array}{ccc} 0 & u & v \\ u^* & 0 & 0 \\ v^* & 0 & 0
\end{array}\right),\qquad L_{-1} \equiv \varphi_2(L_1)= \left(\begin{array}{ccc} 0 & -u & v \\ -u^*
& 0 & 0 \\ v^* & 0 & 0 \end{array}\right)
\end{equation}
and the corresponding system of equations (\ref{eqaniso}) is
\begin{equation} \label{eq3}
\fl \eqalign{
\rmi u_t=u_{xx}-(u(u^*u_x+v^*v_x))_x-4u(|u|^2-|v|^2)+\alpha_1 u\, ,\cr
\rmi v_t=v_{xx}-(v(u^*u_x+v^*v_x))_x-4v(|u|^2-|v|^2)+\alpha_2 v\, ,} \qquad |u|^2+|v|^2=1\, .
\end{equation}
In this case the constant matrix $A_0$ is diagonal $A_0=\mbox{diag}(0,\alpha_1,\alpha_2)$.
Constant solutions of the system (\ref{eq3}) depend on the choice of the constants $\alpha_1,\alpha_2$.
If $\alpha_1=\alpha_2$ and $|\alpha_1|\le 4$ then the constant solution of (\ref{eq3}) is
$$ u=\sqrt{\frac{4+\alpha_1}{8}}e^{i\theta_1},\quad v=\sqrt{\frac{4-\alpha_1}{8}}e^{i\theta_2}
$$
and $\theta_1,\theta_2$ are arbitrary phases. Let us choose $\alpha_1=-\alpha_2=4$. In this case the system (\ref{eq3}) can be written in the form
(\ref{eq:u-v1}), (\ref{eq:u-v2}) and has two constant solutions
\[\eqalign{
\mbox{a)} \qquad  u(x,t) = \rme^{\rmi\theta}, \quad  v(x,t) =0, \cr
\mbox{b)} \qquad  u(x,t) =0, \quad v(x,t) = \rme^{\rmi\theta}.}
\]
where $\theta$ is an arbitrary phase.

\section{Spectral properties of $L$}

The spectral properties  of the Lax operator crucially depend on the choice of the
class of admissible potentials. Below we will consider two different classes
satisfying different boundary conditions:
\begin{equation}\label{eq:}
\eqalign{
\mbox{a)} \qquad \lim_{x\to \pm\infty} u(x,t) =\rme^{\rmi\phi_\pm}, \quad \lim_{x\to \pm\infty} v(x,t) =0, \cr
\mbox{b)} \qquad \lim_{x\to \pm\infty} u(x,t) =0, \quad \lim_{x\to \pm\infty} v(x,t) =\rme^{\rmi\phi_\pm}.}
\end{equation}
This choice of the boundary conditions ensures that the asymptotic potentials $U_\pm(\lambda) = \lim_{x\to\pm\infty}
(\lambda L_1 + \lambda^{-1} L_{-1})$ satisfy
\begin{equation}\label{eq:Upm}
\fl \eqalign{
\mbox{a)} \qquad U_{\pm, \rm as}(\lambda) = \psi_{0,\pm} J_{\rm a}(\lambda) \psi_{0,\pm}^{-1}, \qquad
\psi_{0,\pm} = \frac{1}{\sqrt{2}}\left(\begin{array}{ccc} 1 & 0 & e^{i\phi_\pm} \\ e^{-i\phi_\pm}
& 0 & -1 \\ 0 & -1 & 0 \end{array}\right) , \cr
\qquad J_{\rm a} = (\lambda -\lambda^{-1}) K_1, \qquad K_1=\diag (1,0,-1), \cr
\mbox{b)} \qquad U_{\pm, \rm as} (\lambda) = \psi_{0,\pm} J_{\rm b}(\lambda) \psi_{0,\pm}^{-1}, \qquad
\psi_{0,\pm} = \frac{1}{\sqrt{2}}\left(\begin{array}{ccc} 1 & 0 & -e^{i\phi_\pm} \\ 0
& 1 & 0 \\ e^{-i\phi_\pm} & 0 & 1 \end{array}\right) , \cr
\qquad J_{\rm b} = (\lambda +\lambda^{-1}) K_1, \qquad K_1=\diag (1,0,-1),}
\end{equation}

The Jost solutions are fundamental solutions defined as follows
\begin{equation}
\lim_{x\to\pm\infty}\psi_{\pm}(x,\lambda)  \rme^{-\rmi J(\lambda)x}\psi_{0,\pm}^{-1} =\openone .	
\label{josts}\end{equation}
Due to the existence of reductions the Jost solutions satisfy
the symmetry relations
\begin{eqnarray}
\psi^\dag _{\pm}(x,\lambda^*) &=& (\psi_{\pm}(x,\lambda))^{-1},\label{jostred1}\\
J_1\psi_{\pm}(x,-\lambda)J_1 &=& \psi_{\pm}(x,\lambda),\label{jostred2}\\
J_2\psi_{\pm}(x,1/\lambda)J_2 &=& \psi_{\pm}(x,\lambda).\label{jostred3}	
\end{eqnarray}

Next we introduce the auxiliary functions
\begin{equation}\label{eq:xi-pm}
\eta_{\pm}(x,\lambda)=\psi_{0,\pm}^{-1} \psi_{\pm}(x,\lambda)  \rme^{-\rmi J(\lambda)x},
\end{equation}
$\eta_{\pm}(x,\lambda)$ is solution to the associated system:
\begin{equation}\label{eq:eta}
i\frac{d\eta_\pm} {dx} + U_\pm(x,\lambda) \eta_{\pm}(x,\lambda) -\eta _{\pm}(x,\lambda) J(\lambda)=0, \qquad
\end{equation}
where
\begin{equation}
U_{\pm}(x,\lambda)=\psi_{0,\pm}^{-1}\left(\lambda L_1(x) +\frac{1}{\lambda}L_{-1}(x)\right)\psi_{0,\pm},
\end{equation}
and satisfies the boundary conditions $\lim_{x\to\pm\infty} \eta_{\pm}(x,\lambda) =\openone$.

Equivalently $\eta_\pm(x,\lambda)$ are solutions of the following Volterra-type integral equations:
\begin{equation}
\fl \eta_{\pm}(x,\lambda)=\openone+\rmi\int^{x}_{\pm\infty}\rmd y
\rme^{\rmi J(\lambda)(x-y)}[U_\pm (y,\lambda)-J(\lambda)]\eta_{\pm}(y,\lambda)
\rme^{-\rmi J(\lambda)(x-y)},
\end{equation}

In case a) the Jost solutions are well defined on the real axis in the complex $\lambda$-plane.

Once the Jost solutions are introduced one defines their transition matrix $T(\lambda)$
\begin{equation}
\psi_{-}(x,\lambda)=\psi_{+}(x,\lambda)T(\lambda), \qquad \lambda\in \bbbr.
\label{tmatrix}\end{equation}
As a consequence of symmetries (\ref{jostred1})--(\ref{jostred3}) the scattering matrix
$T(\lambda)$ obeys the following conditions
\begin{eqnarray}
T^\dag (\lambda^*) &=& T^{-1}(\lambda),\label{scatred1}\\
J_1T(-\lambda)J_1 &=& T(\lambda),\label{scatred2}\\
J_2T(1/\lambda)J_2 &=& T(\lambda).\label{scatred3}
\end{eqnarray}
From the Lax representation there follows, that
the scattering matrix  evolves according to the differential equation
\begin{equation}
\rmi\partial_t T+[f(\lambda),T]=0\qquad
\Rightarrow\qquad T(t,\lambda)=\rme^{\rmi f(\lambda)t}T(0,\lambda)
\rme^{-\rmi f(\lambda)t},	
\end{equation}
where
\begin{equation}
f(\lambda)=\lim_{x\to\infty}\sum_{k=-2}^2 \lambda^k A_k(x),
\end{equation}
is the dispersion law of nonlinear equation.

In what follows we will construct the FAS for the special case $\phi_+=\phi_-$; without restriction
we can assume that $\phi_+=\phi_-=0$. Then
$U_{+,\rm as}=U_{-,\rm as}$ and $U_{+}(x,\lambda)=U_{-}(x,\lambda)$.

\subsection{Case a)}
The main tool in constructing the spectral theory of the Lax operator is the
fundamental analytic solution. In the case a) we construct the solutions   $\chi^\pm(x,\lambda)$
which are analytic functions for $\lambda\in\bbbc_\pm $ --  the upper and lower half planes respectively.
In this subsection $J(\lambda)\equiv J_{\rm a)}(\lambda) =(\lambda -\lambda^{-1})K_1$.

First we define $\xi_\pm(x,\lambda)$ as the solutions of the following set of integral equations:
\begin{equation}\label{eq:xip1}
\fl \xi_{kl}^+(x,\lambda)=\delta_{kl}+\rmi\int^{x}_{-\infty}\rmd y
\rme^{\rmi(J_{kk}(\lambda)-J_{ll}(\lambda))(x-y)}\left((U_{-}(y,\lambda)
-J(\lambda))\xi^+(y,\lambda)\right)_{kl}
\end{equation}
for $k\leq l$ and
\begin{equation}\label{eq:xip2}
\fl \xi_{kl}^+(x,\lambda)=\rmi\int^{x}_{\infty}\rmd y
\rme^{\rmi(J_{kk}(\lambda)-J_{ll}(\lambda))(x-y)}\left((U_{-}(y,\lambda)
-J(\lambda))\xi^+(y,\lambda)\right)_{kl}
\end{equation}
for $k>l$. Obviously $\xi^+(x,\lambda)$ is a fundamental solution to eq. (\ref{eq:}). Besides,
due to the appropriate choice of the lower integration limits in eqs. (\ref{eq:xip1}) and
(\ref{eq:xip2}) one finds that the exponential factors in the integrands are falling off
for all $\lambda\in\bbbc_+$. From these basic facts there follows that $\xi^+(x,\lambda)$
is a fundamental analytic solution of eq. (\ref{eq:}) for $\lambda\in\bbbc_+$.

The fundamental analytic solution $\chi^{+}(x,\lambda)$ of the Lax operator $L$ is obtained from $\xi^+(x,\lambda)$ by
applying the simple transformation:
\begin{equation}\label{eq:chi'}
 \chi^{+}(x,\lambda) = \psi_{0,-} \xi^+(x,\lambda) \rme^{\rmi J(\lambda)x}.
\end{equation}

The fundamental analytic solution $\chi^{-}(x,\lambda)$ of the Lax operator $L$ analytic for $\lambda\in\bbbc_-$
is obtained by applying the same transformation:
\begin{equation}\label{eq:chi''}
 \chi^{-}(x,\lambda) = \psi_{0,-} \xi^-(x,\lambda) \rme^{\rmi J(\lambda)x}.
\end{equation}
where $\xi^-(x,\lambda)$ is a solution to the equations
\begin{equation}\label{eq:xim1}
\fl \xi_{kl}^-(x,\lambda)=\delta_{kl}+\rmi\int^{x}_{\infty}\rmd y
\rme^{\rmi(J_{kk}(\lambda)-J_{ll}(\lambda))(x-y)}\left((U_{-}(y,\lambda)
-J(\lambda))\xi^-(y,\lambda)\right)_{kl}
\end{equation}
for $k\leq l$ and
\begin{equation}\label{eq:xim2}
\fl \xi_{kl}^-(x,\lambda)=\rmi\int^{x}_{-\infty}\rmd y
\rme^{\rmi(J_{kk}(\lambda)-J_{ll}(\lambda))(x-y)}\left((U_{-}(y,\lambda)
-J(\lambda))\xi^-(y,\lambda)\right)_{kl}
\end{equation}
for $k>l$.

The fundamental analytic solutions are linearly related to the Jost solutions for $\lambda\in\bbbr$.
These relations are expressed through the factors of Gauss decomposition of $T(t,\lambda)$
\begin{equation}\label{eq:T}
T(t,\lambda)=T^{\mp}D^{\pm}(S^{\pm})^{-1}	
\end{equation}
and have the form:
\begin{equation}\label{eq:rhp'}
\chi^{\pm}(x,\lambda)=\psi_{-}(x,\lambda)S^{\pm}
=\psi_{+}(x,\lambda)T^{\mp}(\lambda)D^{\pm}(\lambda).	
\end{equation}
From the reduction conditions (\ref{scatred1})-(\ref{scatred3}) and eq. (\ref{eq:T}) there follows:
\begin{equation}\label{eq:TGred}
\fl \eqalign{
(S^+(\lambda^*))^\dag = (S^-(\lambda))^{-1}, \quad (D^+(\lambda^*))^\dag = (D^-(\lambda))^{-1}, \quad
(T^+(\lambda^*))^\dag = (T^-(\lambda))^{-1}, \cr
J_1S^\pm (-\lambda)J_1 = S^\pm (\lambda), \quad D^\pm(-\lambda)) = D^\pm(\lambda), \quad
J_1T^\pm (-\lambda)J_1 = T^\pm (\lambda), \cr
J_2S^\pm (1/\lambda)J_2 = S^\pm (\lambda), \quad D^\pm(1/\lambda)) = D^\pm(\lambda), \quad
J_2T^\pm (1/\lambda)J_2 = T^\pm (\lambda), }
\end{equation}
As a consequence there follow reductions on the FAS:
\begin{equation}\label{eq:FAS-red}
\eqalign{
(\chi^{+})^\dag (x,\lambda^*) = \chi^{-}(x,\lambda), \cr J_1\chi^{+}(x,-\lambda)J_1 = \chi^{-}(x,\lambda), \cr
J_2\chi^{\pm}(x,1/\lambda)J_2 = \chi^{\pm}(x,\lambda).}
\end{equation}

From the relation (\ref{eq:rhp'}) one obtains
\begin{equation}\label{riemman}
\fl \chi^{+}(x,\lambda)=\chi^{-}(x,\lambda)G(x,\lambda), \quad G(x,\lambda) =
\rme^{\rmi J(\lambda)x} (S^-)^{-1}S^+(\lambda)\rme^{-\rmi J(\lambda)x}, \quad \lambda\in \bbbr.	
\end{equation}
which can be seen as a  Riemann-Hilbert problem.
Thus the inverse spectral problem can be reduced to a  Riemann-Hilbert
problem to find matrix functions analytic in the upper and lower half
plains of $\lambda$ and satisfying (\ref{riemman}) on the real axis.
This Riemann-Hilbert problem does not allow canonical normalization neither
for $\lambda\to\infty$, nor for $\lambda\to 0$.  The symmetry conditions
replaces (partially) the normalization of the Riemann-Hilbert problem
(compare with \cite{mik_ll}).

\begin{remark}\label{rem:1}
The Riemann-Hilbert problem allows singular solutions as well. The simplest types of
singularities are simple poles and zeroes of the FAS and generically correspond to
discrete eigenvalues of the Lax operator $L$.  Due to the reduction symmetries the
discrete eigenvalues must form orbits of the reduction group. Generic orbits contain octuplets,
so if $\mu_1$ is an eigenvalue, then $-\mu_1$, $\pm \mu^*_1$, $\pm 1/\mu_1$ and $\pm 1/\mu_1^*$
 are eigenvalues too. However,  we can have degenerate orbits. If the eigenvalue $\mu_2 $ lies on the unit circle $|\mu_2|=1$
(resp. if $\mu_3 =-\mu_3^*$ lies on the imaginary axis) we will have quadruplets of eigenvalues. The smallest
degenerate orbit consists of two points only equal to $\pm \rmi$.
\end{remark}

\subsection{Case b)}
\begin{figure}
\centerline{\includegraphics[width=8cm]{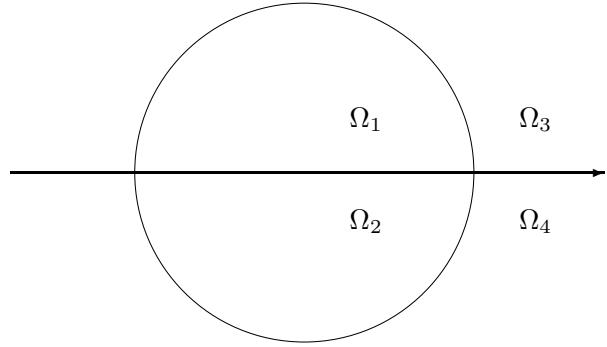}}
\caption{Continuous spectrum of $L$, case b).}\label{spectrum}
\end{figure}

Due to the fact that now $J(\lambda)\equiv J_{\rm b)}$ is proportional to $\lambda + \lambda^{-1}$
we find that the continuous spectrum of $L$ fills up the real axis and the circle with radius $1$
in the complex $\lambda$-plane, see the figure \ref{spectrum}.

The Jost solutions and the scattering matrix
are introduced as in the previous case (see (\ref{josts}) and (\ref{tmatrix})).
Their domain now is the union of the real axis and the unit circle (that is on
the continuous spectrum only). The regions of analyticity are four, denoted
by $\Omega_1$, $\Omega_2$, $\Omega_3$ and $\Omega_4$.

The construction of the Jost solutions is formally possible only for potentials whose $x$-derivatives
are on finite support. Skipping the details we outline the construction of the fundamental analytic solutions in each of the
domains $\Omega_j$, $j=1,\dots ,4$.

The fundamental analytic solutions of eq. (\ref{eq:}) in the domains $\Omega_1\cup \Omega_4$
satisfy the following integral equations
\begin{equation}
\fl \xi^{(s)}_{kl}(x,\lambda)=\delta_{kl}+\rmi\int^{x}_{-\infty}\rmd y
\rme^{\rmi(J_{kk}(\lambda)-J_{ll}(\lambda))(x-y)}\left((U_{-}(y,\lambda)
-J(\lambda))\xi^{(s)}(y,\lambda)\right)_{kl}
\end{equation}
for $k\leq l$ and
\begin{equation}
\fl \xi^{(s)}_{kl}(x,\lambda)=\rmi\int^{x}_{\infty}\rmd y
\rme^{\rmi(J_{kk}(\lambda)-J_{ll}(\lambda))(x-y)}\left((U_{-}(y,\lambda)
-J(\lambda))\xi^{(s)}(y,\lambda)\right)_{kl}
\end{equation}
for $k>l$. In the equations above $s=1$ and $4$ and $\lambda$ is assumed to take values in the domain
$\Omega^{(s)}$.

The corresponding fundamental analytic solutions of  the Lax operator $L$ is obtained from $\xi^{(s)}(x,\lambda)$ by
applying the simple transformation:
\begin{equation}\label{eq:chi's}
 \chi^{(s)}(x,\lambda) = \psi_{0,-} \xi^{(s)}(x,\lambda) \rme^{\rmi J(\lambda)x}.
\end{equation}

The fundamental analytic solution $\chi^{(s')}(x,\lambda)$ of the Lax operator $L$ analytic for $\lambda\in \Omega^{(s')}$
is obtained by applying the same transformation:
\begin{equation}\label{eq:chi''s}
 \chi^{(s')}(x,\lambda) = \psi_{0,-} \xi^{(s')}(x,\lambda) \rme^{\rmi J(\lambda)x}.
\end{equation}
where $\xi^{(s')}(x,\lambda)$ is a solution to the equations
\begin{equation}
\fl \xi^{(s')}_{kl}(x,\lambda)=\delta_{kl}+\rmi\int^{x}_{\infty}\rmd y
\rme^{\rmi(J_{kk}(\lambda)-J_{ll}(\lambda))(x-y)}\left((U_{-}(y,\lambda)
-J(\lambda))\xi^{(s')}(y,\lambda)\right)_{kl}
\end{equation}
for $k\leq l$ and
\begin{equation}
\fl \xi^{(s')}_{kl}(x,\lambda)=\rmi\int^{x}_{-\infty}\rmd y
\rme^{\rmi(J_{kk}(\lambda)-J_{ll}(\lambda))(x-y)}\left((U_{-}(y,\lambda)
-J(\lambda))\xi^{(s')}(y,\lambda)\right)_{kl}
\end{equation}
for $k>l$. In the equations above $s'=2$ and $3$ and $\lambda$ is assumed to take values in the domain
$\Omega^{(s')}$.

As in case a), the fundamental analytic solution are linearly related to the Jost solutions as follows through the Gauss factors
of $T(\lambda)$:
\begin{equation}\label{eq:chi-psi}
\fl \eqalign{
\chi^{(1)}(x,\lambda)  = \psi_-(x,\lambda) S^+(\lambda) =\psi_+(x,\lambda) T^-D^+(\lambda)
, \qquad \lambda \in \bbbr_0\cup \bbbs_+, \cr
\chi^{(2)}(x,\lambda)  = \psi_-(x,\lambda) S^-(\lambda) =\psi_+(x,\lambda) T^+D^-(\lambda)
, \qquad \lambda \in \bbbr_0\cup \bbbs_-, \cr
\chi^{(3)}(x,\lambda)  = \psi_-(x,\lambda) S^-(\lambda)=\psi_+(x,\lambda) T^+D^-(\lambda)
, \qquad \lambda \in \bbbr_1\cup \bbbs_+, \cr
\chi^{(4)}(x,\lambda) = \psi_-(x,\lambda) S^+(\lambda)=\psi_+(x,\lambda) T^-D^+(\lambda)
, \qquad \lambda \in \bbbr_1\cup \bbbs_-,}
\end{equation}
where $S^\pm (\lambda)$, $ T^\pm (\lambda)$ and $D^\pm (\lambda)$ are the Gauss factors of the
scattering matrix (see eq. (\ref{eq:T})) and
\begin{equation}\label{eq:rspm}
\fl \eqalign{
\bbbr_0 &\equiv \{ -1 \leq \re \lambda \leq 1\},  \qquad \qquad
\bbbr_1 \equiv \{ -\infty \leq \re \lambda \leq -1 \} \cup \{ 1 \leq \re \lambda \leq \infty \}, \\
\bbbs_+ &\equiv \{ | \lambda |=1, \; 0\leq \arg \lambda  \leq \pi \} , \qquad
\bbbs_- \equiv \{ | \lambda |=1, \; \pi \leq \arg \lambda  \leq 2\pi \} ,}
\end{equation}

The fundamental analytic solutions of adjacent regions are connected via
\begin{equation}\label{eq:rhp}
\eqalign{
\xi^{(1)}(x,\lambda) =\xi^{(2)}(x,\lambda) G(x,t,\lambda),  \qquad \lambda \in \bbbr_0, \\
\xi^{(4)}(x,\lambda) =\xi^{(3)}(x,\lambda) G(x,t,\lambda),  \qquad \lambda \in \bbbr_1, \\
\xi^{(1)}(x,\lambda) =\xi^{(3)}(x,\lambda) G(x,t,\lambda),  \qquad \lambda \in \bbbs_+, \\
\xi^{(4)}(x,\lambda) =\xi^{(2)}(x,\lambda) G(x,t,\lambda),  \qquad \lambda \in \bbbs_-,}
\end{equation}
where
\begin{equation}\label{eq:Gxt}
G(x,\lambda) = \rme^{\rmi J(\lambda)x} (S^-)^{-1}S^+(\lambda)\rme^{-\rmi J(\lambda)x},
\end{equation}
Thus the inverse spectral problem can be reduced to a generalized
Riemann-Hilbert problem to find piecewise analytic matrix function satisfying conditions (\ref{eq:rhp})
across the contour defined by the continuous spectrum (see Fig. 1).

The reductions imposed on the Jost solutions, $T(\lambda)$ and its Gauss factors  are the same like in eqs.
(\ref{jostred1}) -- (\ref{jostred3}) and (\ref{eq:TGred}). They result in the following relations between the FAS:
\begin{equation}\label{eq:caseb}
\fl \eqalign{
(\chi^{(1)})^\dag (x,\lambda^*) = \chi^{(2)}(x,\lambda), \qquad (\chi^{(4)})^\dag (x,\lambda^*) = \chi^{(3)}(x,\lambda), \cr
J_1\chi^{(1)}(x,-\lambda)J_1 = \chi^{(2)}(x,\lambda), \qquad J_1\chi^{(4)}(x,-\lambda)J_1 = \chi^{(3)}(x,\lambda), \cr
J_2\chi^{(1)}(x,1/\lambda)J_2 = \chi^{(4)}(x,\lambda), \qquad J_2\chi^{(2)}(x,1/\lambda)J_2 = \chi^{(3)}(x,\lambda).}
\end{equation}

\begin{remark}\label{rem:2}

Just like in the previous case (see remark \ref{rem:1} the Riemann-Hilbert problem allows
singular solutions as well. which correspond to discrete eigenvalues of the Lax operator $L$.  So, like
before, generic eigenvalues form octuplets: if $\mu_1$ is an eigenvalue, then $\pm \mu_1$, $\pm \mu^*_1$, $\pm 1/\mu_1$ and $\pm 1/\mu_1^*$
 are eigenvalues too. Since now the unit circle is part of the continuous spectrum of $L$
we can not have discrete eigenvalues on it, but we still can have quadruplets of eigenvalues
on the imaginary axis $\pm \rmi\eta_2$ and $\pm \rmi/\eta_2$, $\eta_2\neq \pm 1$ and real.

\end{remark}

\subsection{Asymptotics of fundamental analytic solution for  $\lambda\to\infty$}

Below we will need the matrix $g_1(x,t)$ which diagonalize  the potential $L_1(x,t)$:
\begin{equation}\label{eq:Lgg}
\fl \eqalign{
L_1 g_1 =  g_1(x,t) K_1,\quad K_1 =  \left(\begin{array}{ccc} 1 & 0 & 0 \\ 0 & 0 & 0 \\
0 & 0 & -1 \end{array}\right) , \cr
g_1 = \frac{1}{\sqrt{2}} \left(\begin{array}{ccc} 1 & 0 & - 1 \\ u^* & \sqrt{2} v & u^* \\
v^* & - \sqrt{2} u & v^* \end{array}\right)  ,
g_1^{-1} = \frac{1}{\sqrt{2}} \left(\begin{array}{ccc} 1 & u & v \\ 0 & \sqrt{2} v* & -\sqrt{2} u^* \cr
-1 &  u & v \end{array}\right) ,}
\end{equation}
\begin{equation}\label{eq:g1g}
a\equiv g_{1,x} g_1^{-1}  =  \left(\begin{array}{ccc} 0 & 0 & 0 \\ 0 & u_x^* u + v_xv^* & u_x^* v - v_x u^*\\
0 & v_x^* u - u_x v^* & v_x^* v + u_x u^* \end{array}\right) .
\end{equation}

For $\lambda \to \infty$ we have to solve the equation
\begin{equation}\label{eq:LL1}
\rmi\frac{\rmd\chi_{\rm as} }{\rmd x} + \lambda L_1 \chi_{\rm as}(x,\lambda) =0
\end{equation}
and determine the asymptotic behavior of $\chi_{\rm as}(x,\lambda)$ for $\lambda \to\infty$.
We introduce:
\begin{equation}\label{eq:chi-t}
\tilde{\chi}_{\rm as}(x,\lambda) = g^{-1}(x) \chi_{\rm as} (x,\lambda) \rme^{-\rmi K_1 \lambda x},
\end{equation}
which satisfy:
\begin{equation}\label{eq:chi_teq}
\rmi \frac{\rmd \tilde{\chi}_{\rm as}} {\rmd x} -\rmi g^{-1}_{1,x}g_1 \tilde{\chi}_{\rm as}(x,\lambda) +
\lambda [K_1,\tilde{\chi}_{\rm as}(x,\lambda)] =0.
\end{equation}
With properly chosen asymptotic conditions $\tilde{\chi}_{\rm as}(x,\lambda)$ will provide the
asymptotics of the fundamental analytic solution. Therefore it will allow an asymptotic expansion of the form:
\begin{equation}\label{eq:chit-as}
\tilde{\chi}_{\rm as}(x,\lambda) = \sum_{s=0}^{\infty} \lambda^{-s} \tilde{\chi}_{k,\rm as}(x) .
\end{equation}
Inserting this expansion into eq. (\ref{eq:chi_teq}) for the first two coefficients of $\tilde{\chi}_{\rm as}(x,\lambda)$ we get:
\begin{equation}\label{eq:chit-k}
\eqalign{
[K_1,\tilde{\chi}_{0,\rm as}(x)] =0 \cr
\rmi \frac{\rmd \tilde{\chi}_{0,\rm as}} {\rmd x} - \rmi g^{-1}_{1,x}g_1 \tilde{\chi}_{0,\rm as}(x) +
[K_1,\tilde{\chi}_{1,\rm as}(x)] =0 , \cr
\rmi \frac{\rmd \tilde{\chi}_{1,\rm as}} {\rmd x} - \rmi g^{-1}_{1,x}g_1 \tilde{\chi}_{1,\rm as}(x) +
[K_1,\tilde{\chi}_{2,\rm as}(x)] =0.}
\end{equation}
Thus we conclude that $\tilde{\chi}_{0,\rm as}(x)$ must be a diagonal matrix of the form:
\begin{equation}\label{eq:chi0}
\fl \tilde{\chi}_{0,\rm as}(x) = \diag \left(\rme^{p(x)},\rme^{-2p(x)},\rme^{p(x)}\right),
\qquad p(x) = \frac{1}{2}\int_{\pm \infty} ^x \rmd y (u^*u_y + v^*v_y),
\end{equation}
and for the off-diagonal part of $\tilde{\chi}^{\rm f}_{1,\rm as}(x)$ we have
\begin{equation}\label{eq:chi1}
\fl \tilde{\chi}^{\rm f}_{1,\rm as}(x) = \frac{\rmi}{2}\left(\begin{array}{ccc}
0 & \sqrt{2}(vu_x-uv_x)\rme^{-p}  & (u^*u_x+v^*v_x)\rme^{p}/2 \\
-\sqrt{2}(u^*v^*_x-v^*u^*_x)\rme^{p} & 0 & -\sqrt{2}(u^*v^*_x-v^*u^*_x)e^{p}\\
-(u^*u_x+V^*v_x)\rme^{p}/2 & \sqrt{2}(vu_x - uv_x)e^{-2p} & 0 \end{array}\right)
\end{equation}
As a result the asymptotic behavior of $\chi(x,\lambda)$ for $\lambda\to\infty$ is given by:
\begin{equation}\label{eq:chi-asm}
\eqalign{
\chi(x,\lambda) \mathop{\simeq}\limits_{\lambda\to\infty} g_1^{-1} \left( \tilde{\chi}_{0,\rm as}(x)
+ \frac{1}{\lambda} \tilde{\chi}_{1,\rm as}(x) + \cdots \right) \rme^{\rmi K_1 \lambda x}.}
\end{equation}
Thus we conclude that the fundamental analytic solution $\chi^\pm$ do not allow canonical normalization for $\lambda\to\infty$.
This difficulty can be overcome by applying a suitable gauge transformation.

The asymptotic behavior of $\tilde{\chi}_{\rm as}(x,\lambda)$ for $\lambda\to 0$ can be derived in a similar way.
One can also use the involution that maps $\lambda$ into $1/\lambda$.

\section{The Wronskian relations and the squared solutions }

Consider
\begin{equation}\label{eq:Wr1}
\fl \eqalign{
\left. \left( i \chi^{-1} B \chi (x,\lambda) - iB\right)\right|_{x=-\infty}^{\infty}  \cr \qquad\qquad  = \int_{x=-\infty}^{\infty} dx\,
\left( \chi^{-1}( \lambda [L_1,B] + \lambda^{-1} [L_{-1},B] + i B_x ) \chi(x,\lambda) - i B_x \right) ,}
\end{equation}
where $B(x,\lambda)$ is for now arbitrary function. We will use below two choices for $B$: the
first one will be $B=J_0$ where $J_0$ is a constant diagonal matrix; then
\begin{equation}\label{eq:Wr2}
\fl \eqalign{
\left. \left( i\chi^{-1} J_0 \chi(x,\lambda) - iJ_0\right)\right|_{x=-\infty}^{\infty}  = \int_{x=-\infty}^{\infty} dx\,
\left( \chi^{-1}( [\lambda L_1 + \lambda^{-1} L_{-1},J_0] ) \chi(x,\lambda)  \right) ,}
\end{equation}
The second choice  is $B(x,\lambda)=\lambda L_1(x) + \lambda^{-1} L_{-1}(x)$ which results in:
\begin{equation}\label{eq:Wr5}
\fl \eqalign{
\left. \left( i \chi^{-1} (\lambda L_1 + \lambda^{-1} L_{-1}) \chi(x,\lambda) \right)\right|_{x=-\infty}^{\infty}\cr
 \qquad \qquad = \int_{x=-\infty}^{\infty} dx\,
\left( \chi^{-1}( i(\lambda L_{1,x} + \lambda^{-1} L_{-1,x}) \chi(x,\lambda)  \right) ,}
\end{equation}

A second class of Wronskian relations contain the variation of the fundamental analytic solutions due to variations of the potentials $L_{\pm 1}$.
\begin{equation}\label{eq:dLpm}
i \frac{\delta\chi }{dx} +(\lambda L_{1} + \lambda^{-1} L_{-1}) \delta\chi(x,\lambda) +
(\lambda \delta L_{1} + \lambda^{-1} \delta L_{-1}) \chi(x,\lambda)=0.
\end{equation}
Thus we obtain:
\begin{equation}\label{eq:Wrd}
\fl \eqalign{
\left. i \chi^{-1} \delta \chi (x,\lambda)\right|_{x=-\infty}^{\infty} = -\int_{x=-\infty}^{\infty} dx\,
\left( \chi^{-1}( (\lambda \delta L_1 + \lambda^{-1}\delta L_{-1})) \chi(x,\lambda) \right) ,}
\end{equation}

The left hand sides of the Wronskian relations are expressed in terms of the scattering data and their variations.
The right hand sides can be viewed as Fourier-type integrals. To make this obvious we take the Killing form of the Wronskian
relations above with the Cartan-Weyl generators.

Let us take the Killing form of eq. (\ref{eq:Wr2}) with one of the Cartan-Weyl generators $E_\alpha$, assume that $J_0=\varphi_2(J_0)$ and use the invariance of the Killing form to get:
\begin{equation}\label{eq:Wr2'}
\eqalign{
\left. \left\langle i \chi^{-1} J_0 \chi (x,\lambda)- iJ_0, E_\alpha\right \rangle\right|_{x=-\infty}^{\infty} \cr
\qquad = \int_{x=-\infty}^{\infty} dx\,
\left\langle \chi^{-1} \left(\lambda [L_1,J_0] + \lambda^{-1} [L_{-1},J_0] \right)\chi(x,\lambda), E_\alpha\right\rangle \cr
\qquad =  \int_{x=-\infty}^{\infty} dx\, \left\langle [L_1,J_0]  , (\lambda e_\alpha (x,\lambda)
 + \lambda^{-1} \varphi_2(e_\alpha)(x,\lambda)) \right\rangle \cr
\qquad =  \int_{x=-\infty}^{\infty} dx\, \left\langle [L_1,J_0]  , \Phi_1(x,\lambda) \right\rangle}
\end{equation}
where
\begin{equation}\label{eq:ssol4}
\Phi_1(x,\lambda) = \lambda e_\alpha(x,\lambda) +  \lambda^{-1} \varphi_2(e_\alpha)(x,\lambda),
\end{equation}
and
\begin{equation}\label{eq:ssol}
e_\alpha(x,\lambda) = \chi^{-1} E_\alpha \chi(x,\lambda).
\end{equation}

Similarly, taking the Killing form of eq. (\ref{eq:Wr5}) with  $E_\alpha$ and using the invariance of the Killing we find:
\begin{equation}\label{eq:Wr5'}
\eqalign{
\left. \left\langle i \chi^{-1} (\lambda L_1 + \lambda^{-1} L_{-1}) \chi (x,\lambda), E_\alpha \right\rangle \right|_{x=-\infty}^{\infty}\cr
 \qquad \qquad = i\int_{x=-\infty}^{\infty} dx\, \left( (\lambda L_{1,x} + \lambda^{-1} L_{-1,x}), e_\alpha (x,\lambda)  \right) ,\cr
 \qquad \qquad  = i\int_{x=-\infty}^{\infty} dx\, \left\langle   L_{1,x} , \Phi_1 (x,\lambda)  \right\rangle .}
\end{equation}

Finally, for the second class of the Wronskian relations we have:
\begin{equation}\label{eq:Wrd2}
\fl \eqalign{
\left. i\left\langle \chi^{-1} \delta \chi (x,\lambda),E_\alpha \right\rangle
\right|_{x=-\infty}^{\infty} &= -\int_{x=-\infty}^{\infty} dx\,
\left\langle  (\lambda \delta L_1 + \lambda^{-1}\delta L_{-1}), e_\alpha (x,\lambda) \right\rangle \cr
&= -\int_{x=-\infty}^{\infty} dx\, \left\langle  \delta L_1 , \Phi_1 (x,\lambda) \right\rangle}
\end{equation}

The Wronskian relations are the main tool in analyzing the mapping between the scattering data and the
potentials $L_{\pm 1}(x)$ of $L$. Indeed, taking $\chi(x,\lambda)$ to be a fundamental analytic solution of $L$ we can express the
left hand sides of eq. (\ref{eq:Wr5'}) (resp. (\ref{eq:Wrd2})) through the Gauss factors $S^\pm$, $T^\mp$
and $D^\pm$ (resp. through the Gauss factors and their variations). The right hand side of eq.
(\ref{eq:Wr5'}) (resp. (\ref{eq:Wrd2})) can be interpreted as a Fourier-like transformation of the
potential $L_1(x)$ (resp. of the variation $\delta L_1(x)$). As a natural generalization of the usual
exponential factors there appear the `squared solutions' $\Phi_1(x,\lambda)$. The `squared solutions'
are analytic functions of $\lambda$ which is important in proving the fact that they form a complete set
of function in the space of allowed potentials of $L$.

\section{Conclusions}

We analyzed the reductions of integrable equations on {\bf A.III}-symmetric spaces and constructed the
FAS for the corresponding Lax operator (\ref{L_3}), (\ref{L1}) with $N=3$. These results can be generalized  for $N>3$,
as well as for  other classes of symmetric spaces, such as {\bf A.II}, {\bf D.III}, {\bf C.I}, {\bf BD.I}.

\section*{Acknowledgements}
We would like to thank V.V.Sokolov for drawing our attention to paper \cite{golsok}.
This work has been supported in part by the Royal Society and the Bulgarian academy of sciences (joint
research project "Reductions of Nonlinear Evolution Equations and
analytic spectral theory"). One of us (T. I. V.) acknowledges support from the European Operational programm HRD,
contract BGO051PO001/07/3.3-02/53 with the Bulgarian Ministry of Education and Science.

\end{document}